# Influence of ferroelastic domain walls on thermal conductivity

P. Limelette[1,±], M. El Kamily[1], H. Aramberri[2], F. Giovannelli[1], M. Royo[3], R. Rurali[3], I. Monot-Laffez[1], J. Íñiguez[2,4], G. F. Nataf[1,*]

[1] GREMAN UMR7347, CNRS, University of Tours, INSA Centre Val de Loire, 37000 Tours, France

[2] Materials Research and Technology Department, Luxembourg Institute of Science and Technology (LIST), L-4362 Esch-sur-Alzette, Luxembourg

[3] Institut de Ciència de Materials de Barcelona, ICMAB-CSIC, Campus UAB, 08193 Bellaterra, Spain

[4] Department of Physics and Materials Science, University of Luxembourg, Belvaux, Luxembourg

± patrice.limelette@univ-tours.fr

* guillaume.nataf@univ-tours.fr

Enabling on-demand control of heat flow is key for the development of next generation of electronic devices, solid-state heat pumps and thermal logic. However, a precise and agile tuning of the relevant microscopic material parameters for adjusting thermal conductivities remains elusive. Here, we study several single crystals of lanthanum aluminate ($LaAlO_3$) with different domain structures and show that ferroelastic domain walls behave as boundaries that act like efficient controllers to govern thermal conductivity. At low temperature (3 K), we demonstrate a fivefold reduction in thermal conductivity induced by domain walls orthogonal to the heat flow and a twofold reduction when they are parallel to the heat flow. Atomistic calculations fully support this experimental observation. By breaking down phonon scattering mechanisms, we also analyze the temperature dependence of the thermal conductivity to derive a quantitative relation between thermal conductivity variations and domain wall organization and density.

# I. INTRODUCTION

Thermal switches, whose thermal conductivity can be tuned by an applied field, are essential elements for preventing degradation in electronic circuits [1], developing environment-friendly refrigeration systems [2–4], achieving high-efficiency thermoelectric devices [5,6], and making novel computing mechanisms utilizing phonons a reality (phononics) [7,8].

Their efficiency is parametrized through the ratio ($R$) of their high ($\kappa_{high}$) and low ($\kappa_{low}$) thermal conductivity states. Some are based on fluids, but solid-state thermal switches operating through conduction mechanisms are more appealing because of their resilience and compactness. For example, a single material such as $VO_2$ exhibits a switching ratio $R\sim1.6$, even though it works in a temperature range limited by the temperature of its transition between thermally insulating and conductive states [9,10]. Phase-change materials exhibit larger thermal conductivity differences depending on their crystallinity [11], but lack a demonstration of reversible switching. Alternatively, multilayers of ferromagnetic and non-magnetic conductive layers can exhibit thermal conductivity ratios $R\sim1.8$ under a magnetic field [12]. However, the main limitation of solid-state thermal switches operating through conduction mechanisms remains their comparatively low thermal conductivity ratios ($R < 1.8$).

Recently, ferroic oxides have been proposed for obtaining fast and reversible solid-state thermal switches with large switching ratios in broad temperature ranges [13–16]. They rely on interactions between phonons conducting heat, and spontaneously occurring topological defects known as domain walls [17] that move in response to an electric field [18] or uniaxial stress [19]. Molecular dynamics simulations reveal that the thermal conductivity could be divided by up to $R=3.7$ under the action of ferroelastic domain walls [14], and first-principles calculations show a thermal conductivity ratio $R=2$ when the density of domain walls is increased with an electric field [15]. Seminal experimental works, where the precise orientation and density of domain walls are unknown, mention the influence of domain walls on the thermal conductivity at low temperature [20–22]. More recently, at room temperature, thermal conductivity variations as high as $R=3$ were demonstrated in ferroelectric thin films with different densities of domain walls [23], and switching ratios as high as $R=1.2$ where obtained under the application of an electric field [24–26].

Nevertheless, some published results question the influence of domain walls on the thermal conductivity, e.g. in bismuth ferrite [27,28]. A reason behind these discrepancies is that most measurements are performed only on thin films at room temperature, with complex domain structures, where the influence of residual strain, substrate and defects can be substantial. Furthermore, the thermal conductivity is assessed by thermoreflectance, after removing the response from the interface with the substrate that dominates the initial signal, and only in one direction, missing out anisotropies induced by different orientations of domain walls. These anisotropies are also excluded from current models relying on an interfacial thermal resistance at the wall [23,27]. As such, questions of how domain walls influence the thermal conductivity and how their organization and density should be tuned to have a large impact are still fully open.

In this paper, we investigate the thermal conductivity of bulk single crystals of ferroelastic lanthanum aluminate (LaAlO$_3$) with different density and orientation of domain walls. Here, we measure the thermal conductivity of five samples in different directions and in a broad temperature range – from room temperature down to 2 K. In this temperature range, i.e. far from the phase transition occurring near 850 K, domain walls in LaAlO$_3$ have an estimated thickness of 20 Å [29] and are pinned [30,31], which means that the domain structure is stable and does not change. This unique set of data reveals large thermal conductivity variations below 10 K when the density of domain walls orthogonal to the heat flow is varied, leading to a fivefold reduction in thermal conductivity. It also reveals that domain walls have an influence on heat flows even if they are parallel to them. Still, the thermal conductivity of the material depends of the direction of the heat flow with respect to domain walls, with a twofold difference in thermal conductivity between both directions. We account for these results by analysing the full temperature dependence of the measured thermal conductivity within the framework of the Holland model by considering all relevant phonon scattering mechanisms. We are thus able to relate quantitatively the evolution of the thermal conductivity to ferroelastic domain walls densities and orientations, successfully described by the Casimir limit, here calculated numerically to account for finite size effects. The influence of domain walls on the thermal conductivity is also computed using atomistic simulations of the heat transport (both parallel and orthogonal to the walls) based on second-principles models. By demonstrating that heat flows can be effectively controlled by domain walls, these results provide a general approach to model the influence of these planar defects on thermal conductivity, and pave the way for designing efficient, microscopic thermal switches.

## II. RESULTS
### A. Ferroelastic domains engineering

Figure 1 shows images of the 5 commercial single crystals of LaAlO$_3$ studied, whose surfaces are orthogonal to the [001]$_{pc}$ direction ("pc" stands for pseudo-cubic), obtained with an optical microscope working in reflection (images in transmission in Supplementary Note 1 [32]). Sample F [Fig. 1(a)] has been annealed in air at 1200 K for 6 h with a heating and cooling rate of ~10 K min$^{-1}$ to obtain a monodomain state, i.e. free of domain walls, which can be achieved thanks to a redistribution of defects at high temperature that act as pinning centres for domain walls [33]. Since there are no domain walls, the sample appears as uniform both in reflection and transmission [Fig. 1(a) and Fig. S1]. Note that faint structures optically visible are not ferroelastic domains but negative traces of previous domain patterns that remain as a surface relief [34]. Fig. 1(b) (sample A) and 1(c) (sample C) are as-received single crystals with a very regular domain pattern, clearly visible because of differences in reflectivity between ferroelastic domains. Domain walls are strictly orthogonal to the surface, indicating that they are planes orthogonal to the $[010]_{pc}$ direction, in agreement with the literature [35,36]. Domain structures observed in Fig. 1(e) (sample B) and Fig. 1(f) (sample D) result from quenches of single crystals from 680 K to a room temperature silicon-oil/water mixture implying a large increase in the number of domain walls, which remain mostly orthogonal to the $[010]_{pc}$ direction. For all samples, the direction of the ferroelastic distortion in the domains is indicated in Fig. 1(d). The size of the domains, and hence the density of domain walls, has been measured directly from optical images and is shown in Fig. 1(g) for all samples. Samples A and C exhibit analogous distributions, with a mean sizes around 19 μm +/- 2 μm and 15 μm +/- 2 μm, respectively. Samples B and D exhibit smaller domains, and thus higher density of domain walls, with a mean size around 9 μm +/- 2 μm. The number of domains is smaller in samples A and B, whose long lengths are along the $[100]_{pc}$ direction, compared to samples C and D, but the volume occupied by domain walls remains identical.

### B. Thermal measurements

The temperature dependence of the measured thermal conductivity κ is shown in Fig. 2(a). It is probed in directions where the ferroelastic distortion of domains is the same by symmetry [Fig. 1(d)], and thus cannot influence the thermal conductivity. The three regimes of transport usually observed in insulating oxides are recovered: the low temperature boundary scattering region characterized by a power law *T*-dependence, the presence of a high maximum near 25 K that depends on the effects of impurities, followed by a decrease at higher temperatures due to Umklapp scattering. At room temperature, thermal conductivity values lie between 11 and 13 W m$^{-1}$ K$^{-1}$, in agreement with literature [21]. In the boundary scattering regime, a decrease of the thermal conductivity is observed in samples with domain walls parallel to the heat flow (A and B), which is even stronger when domain walls are orthogonal to the flow (C and D). This is illustrated in Fig. 2(b) by the strong enhancement of the switching ratio *R* achieving values as high as 5 near 2 K and demonstrate the effect of domain walls on thermal transport. We verified that annealing and quenching do not influence the thermal conductivity if the samples stay monodomain (Supplementary Note 3 [32]). In particular, in the optical images of quenched samples some cracks are visible (Supplementary Note 1 and 3 [32]). Even though these cracks are mostly at the surface and do not extend across the thickness of the samples, except on the edges, they are particularly visible because the images are taken in transmission. We verified that these cracks have little influence on the thermal conductivity by comparing in Fig. S2 [32] the thermal conductivity of two monodomain samples: one obtained after annealing (sample F) and one obtained after quenching. At 3 K, the thermal conductivity of the quenched sample is reduced by a factor 1.4 only, far from the fivefold reduction in thermal conductivity observed between samples F and D.

To further analyze the influence of domain walls, we performed specific heat measurements down to 2 K [Fig. 2(c)]. Below 10 K the usual low temperature Debye regime is recovered with a well-defined $T^3$ behavior, characteristic of acoustic phonons contribution. We calculate the Debye temperature $T_D$=375 K according to the expected contribution $C(T<<T_D)=12\pi^4/5\ N_{Av}\ k_B\ (T/T_D)^3$, with $N_{Av}$ the Avogadro's number and $k_B$ the Boltzmann constant. By using the mass density $M/V$=6.52 × $10^6$ g m$^{-3}$ the molar mass $M_{mol}$= 213.88 g mol$^{-1}$, and $\frac{N}{V} = \frac{M}{V}\frac{N_{Av}}{M_{mol}}$, the sound velocity is deduced such that $v = \frac{k_B T_D}{\hbar}\left(6\pi^2 \frac{N}{V}\right)^{-1/3} \approx 4773\ m\ s^{-1}$, in agreement with literature [21,37] ($\hbar = h/2\pi$ is the Planck constant). The temperature dependence of the phonon mean free path $L$, calculated as $\kappa = \frac{1}{3}C_{vol}vL$, is unveiled in Fig. 2(d). In the low temperature boundary scattering regime it saturates differently depending on the orientation and density of domain walls. In particular, $L$ is more constrained when domain walls are orthogonal to the heat flow and their density is high, as in sample D compared to sample C, rather than in the case of domain walls parallel to the heat flow as in samples A and B. In the latter case, it is recovered that a higher domain-wall density in sample B reduces more efficiently the mean free path than in sample A. Whichever the orientation of domain walls, $L$ is always lower than in the monodomain sample F.

### III. DISCUSSION
#### A. Simulations of heat transport with a domain wall

We first compare our results with atomistic heat transport simulations. To this end, we derive a second-principles potential for LaAlO$_3$ (see Appendix B) and we use it to optimize cells with domain walls along the {001}$_{pc}$ direction, which we find to be ~25 Å wide in the low temperature limit, close to the experimental value of 20 Å (ref. [29]). We then make use of non-equilibrium Green's function techniques to compute the thermal conductance of LaAlO$_3$ in three configurations: monodomain state, with domain walls orthogonal and parallel to the heat flow [Figs. 3(a) and 3(b)]. We observe that domain walls reduce the thermal transport moderately when parallel to the heat flow and do so more markedly when orthogonal to it. This agrees very well with our experimental results.

In the simulation, the fact that the difference in thermal conductances increases with temperature indicates that the thermal resistance is attributed mostly to optical modes. This is corroborated by the simulated frequency-resolved transmission curves [Figs. 3(c) and 3(d)], which show that optical modes of low and medium frequencies (3-20 THz) undergo a stronger scattering (and the scattering is more pronounced in the orthogonal configuration). In contrast, the experiments show that acoustic modes dominate at low temperature. The discrepancy may arise from the fact that the real (experimental) walls in the multidomain configuration induce a local strain, especially in the high wall-density limit, while in the simulations we considered a relatively sparse regime. This strain could induce large scattering in the acoustic modes, and hence reduce further the thermal conductivity. Also, the experiments could be measuring partly additive scattering by consecutive walls if the scattered phonons recover some of their population through anharmonic interactions before hitting the next wall. This effect would not be captured by the simulations since they are carried out in the harmonic approximation and do not account for phonon-phonon scattering. For that same reason, the simulated thermal transport curves increase monotonically as a function of temperature, as phonons become increasingly populated. Including anharmonic scattering would recover the correct high-temperature behavior, where the effect of the domain walls vanishes because phonon-phonon scattering becomes dominant but comes at a prohibitive computational cost given the sizes of the simulation boxes employed. Still, our simulations give the right trends for the thermal conductivity reduction caused by domain walls in $LaAlO_3$.

## B. Thermal conductivity of the monodomain sample

In a more general framework, the thermal conductivity can be described by using the semi-classical Boltzmann equation in the frame of the relaxation time approximation $\tau$ for phonons characterized by Bose-Einstein statistics with zero chemical potential. It is written as the integral over the acoustic phonon frequencies by using a Debye model [38]:

$$\kappa = \frac{k_B}{2\pi^2 v} \left(\frac{k_B T}{\hbar}\right)^3 \int_0^{x_D} \tau \frac{x^4 e^x}{(e^x - 1)^2} dx \qquad (1)$$

Here, frequencies are made dimensionless by introducing variables $x = \hbar\omega/k_B T$ and $x_D = \hbar\omega/k_B T_D$. The relaxation time depends on phonon scattering mechanisms and is in most cases a function of frequency and temperature such as $\tau = \tau_T \left(\frac{k_B T}{\hbar \omega_0}\right)^\theta x^\theta$ with a possibly $T$-dependent $\tau_T$ and an exponent θ characteristic of the considered scattering process, $\omega_0$ being a related characteristic frequency. At low temperature (typically $T < T_D/20$), the previous integral can be extended to infinity and in the simple case of one type of scattering, the thermal conductivity can be calculated analytically by introducing Gamma Γ and Riemann zeta ζ functions:

$$\kappa \approx \frac{k_B}{2\pi^2 v} \left(\frac{k_B T}{\hbar}\right)^{3+\theta} \frac{\tau_T}{\omega_0^\theta} \int_0^\infty \frac{x^{4+\theta} e^x}{(e^x-1)^2} dx = \frac{k_B}{2\pi^2 v} \left(\frac{k_B T}{\hbar}\right)^{3+\theta} \frac{\tau_T}{\omega_0^\theta} \zeta(4+\theta)\Gamma(5+\theta) \qquad (2)$$

It follows in the case of a boundary scattering, for which the relaxation time is neither frequency nor temperature dependent (θ=0 and $\tau = \tau_B$), that the thermal conductivity varies as $T^3$ and $\kappa = \frac{1}{3} C_{vol} v L_B$ with the temperature independent mean free path $L_B = v\tau_B$.

For the monodomain sample of LaAlO$_3$, this regime is never reached since the measured thermal conductivity varies as $T^2$ at low temperature in Fig. 2(a). Already observed in some perovskite oxides [20], this behavior is the signature of phonon scattering by dislocations [39] with a linear frequency dependence of τ$^{-1}$, namely with θ = -1 and no temperature dependence of the relaxation time. Therefore, it follows from Eq. (2) that κ varies as $T^2$ at low temperature, which explains why the mean free path in Fig. 2(d) is not constant. Within the model of Holland [40], which considers two types of phonon polarization with 1 longitudinal (L) and 2 transverse (T) modes, $\kappa = \frac{1}{3}\kappa_L + \frac{2}{3}\kappa_T$ where $\kappa_L$ and $\kappa_T$ are both defined as in Eq. (1) and the factors 1/3 and 2/3 originate from the numbers of longitudinal and transverse modes, respectively. Additional scattering processes are also considered within the model by including crystalline boundaries, impurities through mass differences, the three-phonon Normal processes (L and T) as well as the Umklapp ones (L and T), and dislocations, such that the relaxation time entering Eq. (1) is $\tau^{-1} = \sum_j \tau_j^{-1}$ where each $\tau_j$ is the relaxation time for a single scattering mechanism [40].

The transverse contribution is further decomposed as $\kappa_T = \kappa_{TN} + \kappa_{TU}$ in order to take into account the specific frequency range of transverse Umklapp scattering [40], which has been adjusted numerically to fit the thermal conductivity. The best adjustment is reached here with a transverse cutoff frequency $\omega_D/1.5$ as summarized in Table 1. As shown in Fig. 4(a), the thermal conductivity measured in the monodomain sample F is successfully described by these 3 contributions, L, TN and TU. According to their definition in Table 1, the total relaxation times used for the calculations are:

$$\tau_L^{-1} = \tau_B^{-1} + \tau_I^{-1} + \tau_D^{-1} + \tau_{NL}^{-1} + \tau_{UL}^{-1} \quad , \text{for } 0 < \omega < \omega_D$$
$$\tau_{TN}^{-1} = \tau_B^{-1} + \tau_I^{-1} + \tau_D^{-1} + \tau_{NT}^{-1} \quad , \text{for } 0 < \omega < \omega_D/1.5$$
$$\tau_{TU}^{-1} = \tau_B^{-1} + \tau_I^{-1} + \tau_D^{-1} + \tau_{UT}^{-1} \quad , \text{for } \omega_D/1.5 < \omega < \omega_D$$

### C. Thermal conductivity of multidomains samples

We have refined the parameters listed by calculating numerically the integrals defined in Eq. (1) at each temperature and for each component to fit the measured thermal conductivity in all the samples. The found factors $A_i$ (Table 1) agree quantitatively with the ones reported for several insulating oxides [41]. One must also emphasize that the only parameters that have been varied from one sample to the other are the dislocation factor and the boundary length $L_B$, which depends on the density and orientation of ferroelastic domains in the samples. Since dislocations act as nucleation or pinning centers for domain walls, one may assume that the contribution from the walls likely takes over the contribution from dislocations. This could then explain why it has been found that the dislocation factor decreases from $2.6 \times 10^{-5}$ (sample F) down to $1.15 \times 10^{-5}$ (samples A and C) and 0 (samples B and D) when the domain-wall density increases. While it remains difficult to ascribe a precise dislocation density from the $A_D$ factor [42], it is known to be proportional to the dislocation density [42,43], and similar values have been found in materials with dislocation densities ranging from $1 \times 10^7$ up to $5 \times 10^7$ cm$^{-2}$, which seems here a reasonable order of magnitude.

### D. Specularity parameter

Boundary scattering has been described here within the framework of Casimir [44,45]. It assumes that the temperature is so low that phonons interact only with the boundary and that the scattering is completely diffuse. This implies that incident phonons which are absorbed by the various surfaces of the sample are then re-emitted with the equilibrium distribution corresponding to the local temperature. However, it appears in Fig. 2(d) that the mean free paths still increase at low temperatures in contrast to the expected constant boundary regime behavior. This is because the boundary scattering of phonons as described by Casimir assumes perfectly diffusive surfaces where phonons are scattered with equal probability into any directions, which is only ensured if the roughness of the surface is higher than the phonon wavelength. Nevertheless, the latter increases typically as $\lambda \approx \frac{hv}{2.821 k_B T}$ when the temperature is lowered, and the previous diffusive criterion is inevitably no longer valid for a sufficiently smooth surface, as shown in Fig. 4(b). In this case, phonons undergo specular reflection on the boundary rather than diffuse scattering and their effective mean free path increases beyond the boundary length usually referred to as the Casimir limit. This effect has been described by Ziman who has proposed to account for it through an average specularity parameter $p$ dependent on the roughness ($\eta$) distribution $P(\eta)$ and the phonon wavelength $\lambda$ (ref. [46]):

$$p = \int P(\eta) e^{\frac{-16\pi^2 \eta^2}{\lambda^2}} d\eta \approx \int_0^{\lambda/4\pi} P(\eta) d\eta \text{ and } L_B^{\text{eff}} = \frac{1+p}{1-p} L_B \qquad (3)$$

Accordingly, an effective mean free path $L_B^{\text{eff}}$ is related to the boundary length $L_B$ in Eq. (3) by considering multiple specular reflections, which lead to a strong enhancement if $p \to 1$ (perfectly smooth surface) or to the bare boundary length if $p = 0$ (diffuse scattering). Whereas the roughness distribution is *a priori* unknown, one may assume that for a polished surface this is a strongly decreasing function of $\eta$ which can be approximated by an exponential form such as $P(\eta) \approx \frac{e^{-\eta/\eta_0}}{\eta_0}$ (ref. [47]). This specularity parameter becomes then $p \approx 1 - e^{-\lambda/4\pi\eta_0}$ and $L_B^{\text{eff}} = (2e^{\lambda/4\pi\eta_0} - 1)L_B$, by considering the dominant phonon wavelength $\lambda \approx \frac{hv}{2.821 k_B T}$. As an example, a comparison is made in Fig. 4(c) between the mean free path inferred from thermal conductivity measurements in sample D without specularity, and the expected ones for a boundary length of 15 μm with a rms roughness $\eta_0 = 7$ nm. The effect of specular reflections appears below ~10 K which corresponds to the expected crossover in Fig. 4(b) between the high temperature diffuse scattering regime (when $\lambda < \eta_0$), and the low temperature specular regime (when $\lambda > \eta_0$). It follows that instead of saturating up to the boundary length, the mean free path still increases if temperature decreases [Fig. 4(c)]. Another comparison is also made by considering a frequency-dependent specularity parameter [48] (as explained in Appendix E) to take into account the different effects of the roughness depending on the short or long phonon wavelengths. Shorter and likely more realistic boundary lengths $L'_B$ are then inferred (Table 2).

### E. Casimir limits from domain distributions

To discuss quantitatively the role of the density and orientation of domain walls, the optically measured sizes of domains must be converted into a relevant boundary length that can be precisely compared to the boundary length $L_B$ (or $L_B$') inferred from the analysis of the thermal conductivity. For this, as explained in Appendix F, we have calculated the so-called Casimir limit which accounts for the finite size effects in the samples (e.g. size of ferroelastic domains) as a function of their length $c$ in Fig. 4(d), by using their experimental dimensions (width $a$=1600 μm, thickness $b$=500 μm).

For small rectangular rods of extension less than 100 μm it appears that the Casimir limit is mainly given by one configuration factor (see Appendix F) such as $3F_{zz}/2\pi$ according to Eq. (A6), due to the smallness of the lateral surfaces. This factor will thus be used to analyze the effect of domain walls orthogonal to the heat flow. The variation of the Casimir limit as a function of the width $a$ in the inset helps to understand the effect of domain walls parallel to the flow because in this case the limit of infinite rod appears justified. In fact, we find that each domain can be considered as an independent rectangular rod with its own thermal conductivity. If walls are orthogonal to the flow, the sample can be represented by a collection of domains in series from the transport point of view, whereas if walls are parallel to the flow, domains must be considered in parallel. It follows that in the former case thermal resistances are additive whereas in the latter thermal conductances are additive. By writing the total thermal conductivity in the boundary regime as $\kappa_{\perp,\parallel} = \frac{1}{3} C_{vol} v L_{C\perp,\parallel}$, depending on the walls' orientation (orthogonal or parallel), with the domain sizes $c_i$ or $a_i$ respectively, the macroscopic Casimir limit $L_{C\perp,\parallel}$ is then related to the distribution of domains through their size and respective Casimir limit $L_{C,i}$:

$$\begin{array}{l}\left(\kappa_\perp \frac{ab}{c}\right)^{-1} = \sum_i \left(\kappa_i \frac{ab}{c_i}\right)^{-1} \\ \kappa_\parallel \frac{ab}{c} = \sum_i \kappa_i \frac{a_i b}{c}\end{array} \Rightarrow \begin{array}{l} L_{C\perp} = \frac{c}{\sum_i c_i/L_{C,i}} \\ L_{C\parallel} = \sum_i \frac{a_i}{a} L_{C,i}\end{array} \quad (4)$$

The resulting macroscopic Casimir limits $L_C$ are summarized in Table 2 and compared to the boundary lengths $L_B$ (and $L_B$') found experimentally from the fit in Fig. 2(d).

### IV. CONCLUSIONS

Thermal conductivities of the monodomain sample and the samples with domain walls orthogonal or parallel to the heat flow [Fig. 2(a)] are successfully reproduced in the frame of the Holland model, by using boundary lengths indicated in the caption with a specular enhancement factor with a rms roughness $\eta_0$=7 nm (or 4 nm for the frequency-dependent specularity). Accordingly, the mean free path of the 5 samples appears suitably described in Fig. 2(d) with boundary lengths originating from different domain sizes in agreement with the inferred Casimir limits for samples F, A and B ($L_B \approx L_C$) or consistent with $L_B \approx L_C \times (1.7 \pm 0.1)$ for samples C and D. These results demonstrate that domain walls behave as boundaries that act as thermal conductivity controllers according to their orientation. The distribution of domain walls implies a distribution of thermal conductivity considered either in series or in parallel depending on the orientation of the walls with respect to the heat flow. The fact that the found boundary length in the orthogonal configuration gets closer to the Casimir limit if one considers a frequency-dependent specularity parameter likely suggests that a more precise description of domain walls, including roughness distributions, is required in these conditions of very short domain sizes.

Overall, we demonstrated that tuning the number and orientation of ferroelastic domain walls is a powerful way to govern thermal conductivity, and we provided a general approach to model the influence of these planar defects on thermal conductivity, paving the way for efficient and compact thermal switches in ferroic materials.

## ACKNOWLEDGMENTS

This paper was cofunded by the European Union [European Research Council (ERC), DYNAM-HEAT, No. 101077402]. Views and opinions expressed are, however, those of the authors only and do not necessarily reflect those of the European Union or the ERC. Neither the European Union nor the granting authority can be held responsible for them. Work at LIST was supported by the Luxembourg National Research Fund through Project No. FNR/C21/MS/15799044/FERRODYNAMICS. We acknowledge financial support by MCIN/AEI/10.13039/501100011033 under grant PID2020-119777GB-I00, and the Severo Ochoa Centres of Excellence Program under grant CEX2019-000917-S, and by the Generalitat de Catalunya under grant 2021 SGR 01519. Low-temperature measurements exploring the effect of domain walls on thermal conductivity in LaAlO3 were also made concurrently at Belfast by Bogdan Zhigulin, Fran Kurnia, Marty Gregg, and Raymond McQuaid. This led to initiation of the second-principles modeling study by HA, MR, RR, and JI.

## APPENDIX A: DOMAIN ENGINEERING

Single crystals of LaAlO$_3$ 10x10x0.5 mm were obtained from Pi-Kem Ltd, with surfaces orthogonal to the [001]$_{pc}$ direction and edges along {001}$_{pc}$ directions. They were cut with a diamond wire saw (Well 3500 Precision) in smaller pieces (~1.6x8x0.5mm), keeping edges along {001}$_{pc}$ directions. Their main surfaces (~1.6x8mm) were optically polished by the supplier, leading to RMS roughness below 1 nm (measured by atomic force microscopy). Samples F, A, C and D come from the same single crystal of 10x10x0.5 mm.

At room temperature, LaAlO$_3$ exhibits a rhombohedral structure, with space group $R\bar{3}c$ (ref. [49]). It thus exhibits eight domain states: four related to the loss of point-group symmetry and four related to the loss of translational symmetry (antiphase boundaries) [34,50,51]. The former leads to six possible pairs of domains. For each of the six pairs there are two possible orientations of domain walls (Supplementary Note 2 [32]). The domain structure was observed with an optical microscope (Olympus BX60) operating in reflection with a polarizer and an analyser, and a x10 objective.

Quenches of single crystals from 680 K to a room temperature silicon-oil (Fisher Scientific)/distilled water mixture (~10/90 volume ratio) came after several attempts at different temperatures with other quenching mediums. Typically, quenching in air and liquid nitrogen did not affect the domain structure, contrary to quenching in oil and water. However, the latter led to the formation of significant cracks in the samples, hence the choice of an oil/water mixture as the best option to reach high density of domain walls without breaking samples.

## APPENDIX B: NUMERICAL SIMULATION OF THE THERMAL CONDUCTANCE REDUCTION IN THE HARMONIC APPROXIMATION

We derive a second-principles [52,53] model for LaAlO$_3$ based on first-principles simulations. The density functional theory calculations used to fit the data were carried out using the Vienna Ab initio Simulation package (VASP) [54,55], making use of the PBEsol [56] implementation of the generalized gradient approximation for the exchange-correlation functional. The model can be viewed as a multi-variable Taylor series of the potential energy of LaAlO$_3$, taking the perovskite cubic phase as the reference structure and treating all atomic distortions explicitly; we work with an effective potential that includes harmonic and anharmonic interatomic couplings up to fourth order.

Domain walls were optimized by seeding an abrupt domain wall in simulation cells elongated in the direction perpendicular to the domain wall and performing Monte Carlo simulated annealings starting from 10 K, which yield smooth profiles for the octahedra rotation pattern. Two domain walls are included in the simulation box to conform with periodic boundary conditions.

The thermal conductance was calculated using non-equilibrium Green's function techniques as described in ref. [57], where the second-order interatomic force constants (IFCs) were computed with the help of the phonopy package [58]. For the configuration with domain walls orthogonal (respectively parallel) to the heat transport (taken as the $[100]_{pc}$ direction), a simulation box of 28x2x2 (respectively 3x14x2) pseudocubic 40-atom unit cells was used to compute the IFCs. The latter were Fourier-transformed in **k** space over the $[010]_{pc}$ and $[001]_{pc}$ directions, and the thermal conductance along $[100]_{pc}$ was evaluated with a supercell of 28x1x1 (respectively 3x14x1) over a grid of 9x9 (respectively 1x15) transverse **k** points. The contacts were defined at both extrema of the supercell with symmetric regions made of 2x1x1 (respectively 1x14x1) pseudocubic cells.

**APPENDIX C: SPECIFIC HEAT AND THERMAL CONDUCTIVITY MEASUREMENTS**

The specific heat has been determined with a calorimeter of a Physical Properties Measurements System from Quantum Design using a relaxation method with a temperature rise of 2% of the sample temperature. Each measurement has been duplicated without the sample to compensate for the temperature dependence of the specific heat of the grease used for a good thermal contact between the sample and the calorimeter platform. The parameter indicating the quality of the thermal contact between the sample and the platform, the so-called sample coupling, has remained between 98% and 100% during each measurement, ensuring thus the reliability of the results. The sample measured was a small piece of 1.7x1.7x0.5 mm cut from the same single crystal as sample B.

To perform thermal conductivity measurements, 100 nm gold electrodes have been deposited through a mask to ensure low contact resistances by using silver paste. Thermal conductivity measurements have been performed under a maintained secondary vacuum ($P < 10^{-5}$ mbar) by using a four points configuration with a Physical Properties Measurements System from Quantum Design and temperatures rise of 1%. All the measurements have been performed at decreasing and then increasing temperatures to avoid thermal hysteresis by averaging both measurements.

**APPENDIX D: DEBYE MODEL OF THE SPECIFIC HEAT**

We emphasize that the Debye model here used accounts for the 3 expected acoustic phonon modes, 1 longitudinal and 2 transverse, which lead (as expected from a physical point of view) to a high temperature asymptotic specific heat $C(T \gg T_D) = 3\,N_{av}k_B$ of the order of 25 J K$^{-1}$ mol$^{-1}$, the rest being ascribed to the 12 optical phonon modes, which do not contribute at low temperatures. This explains the difference between the inferred Debye temperature here and those that can be found in literature where the authors frequently use an effective temperature-dependent Debye temperature to account for the overall variation of the specific heat. Then, their values need to be divided by $5^{1/3}$ due to the 5 atoms in the unit formula LaAlO$_3$ to recover the Debye temperature found here. Both models lead to the same value when this extra term is compensated for in the calculation of the sound velocity. The experimental determination of $T_D$ allows to compute numerically the full temperature dependence of the Debye contribution as defined below.

$$C_{\text{Debye}} = 9N_{Av}k_B \left(\frac{T}{T_D}\right)^3 \int_0^{T_D/T} \frac{x^4 e^x}{(e^x-1)^2} dx \qquad (A1)$$

Therefore, the knowledge of the specific heat and the sound velocity allows to infer the mean free path from the thermal conductivity.

## APPENDIX E: FREQUENCY DEPENDENT SPECULARITY PARAMETER

Instead of using a constant specularity parameter as explained in the main text, the introduced formalism can be extended to account for its frequency dependence by using the same roughness distribution $P(\eta) \approx \frac{e^{-\eta/\eta_0}}{\eta_0}$ without the approximation for Eq. (3) as below:

$$p(\lambda) = \int_0^\infty \frac{e^{-\eta/\eta_0}}{\eta_0} e^{\frac{-16\pi^2 \eta^2}{\lambda^2}} d\eta = \frac{\lambda}{8\sqrt{\pi}\eta_0} e^{\left(\frac{\lambda}{8\pi\eta_0}\right)^2} Erfc\left(\frac{\lambda}{8\pi\eta_0}\right) \text{ with } Erfc(x) = \frac{2}{\sqrt{\pi}} \int_x^\infty e^{-u^2} du \quad (A2)$$

By relating the wavelength to the frequency as $\lambda = \frac{2\pi v}{\omega}$, one can therefore introduce a frequency-dependent specularity parameter $p(\omega)$ in the effective boundary length $L_B^{\text{eff}}(\omega) = \frac{1+p(\omega)}{1-p(\omega)} L_B'$ entering in Eq. (1) which is then integrated numerically.

## APPENDIX F: CALCULATION OF THE CASIMIR LIMIT FROM DOMAIN DISTRIBUTIONS

The Casimir regime assumes that incident phonons which are absorbed by the various surfaces of the sample are then re-emitted with the equilibrium distribution corresponding to the local temperature. Since the emitted energy varies as $T^4$, the temperature can be expanded as $T^4(z) = T^4(0) + 4T^3(0) z (dT/dz)$ by assuming a constant temperature gradient in the $z$ direction, with $z=0$ the center of the sample. By dividing the emitted energy flux by the cross-section $S$ and $(dT/dz)$, one can then write the thermal conductivity $\kappa_C$ in the Casimir limit regime.

$$\kappa_C = \frac{1}{S} 4T^3 \frac{\pi^4 k_B^4}{15 h^3}\left(\frac{3}{v^2}\right) \sum_i \iint \frac{z\cos\theta_{zi}\cos\theta_{xy}}{r_{zi}^2} dS_{zi} dS_{xy} = \frac{\pi}{10} \frac{k_B}{v^2} \left(\frac{k_B T}{\hbar}\right)^3 \sum_i F_{zi} \qquad (A3)$$

Here, $dS_{xy}$ is an element of the cross section and the sum runs over all the surfaces of the sample by involving the corresponding configuration factor $F_{zi} = 1/S \iint \cos\theta_{zi}\cos\theta_{xy} z/r_{zi}^2 \, dS_{zi} dS_{xy}$ with $r_{zi}$ the distance between $dS_{xy}$ and $dS_{zi}$, $\theta_{xy}$ and $\theta_{zi}$ the angles between $r_{zi}$ and the normal vectors to the elements $dS_{xy}$ and $dS_{zi}$ (ref. [43–45]). Also, it is assumed that the sound velocity $v$ is isotropic. It follows then that the classical relation between the thermal conductivity and the mean free path in the Casimir regime is recovered by relating the latter to the configuration factors $F_{zi}$.

$$\kappa_C = \frac{1}{4\pi} \frac{C_D}{V} v \sum_i F_{zi} = \frac{1}{3} \frac{C_D}{V} v L_C \Rightarrow L_C = \frac{3}{4\pi} \sum_i F_{zi} \qquad (A4)$$

In the case of an infinite circular cylinder, $L_C$ is equal to its diameter. If it is an infinite square rod $L_C$ is about 1.12 times the side of the square [44,45], and Harrison and Pendrys [59] have provided an analytical expression in the case of an infinite rectangular rod of thickness $b$ and width $a$ which is slightly simplified below, with the ratio $r=b/a$.

$$L_C^\infty = \frac{3}{4\pi}\left(2F_{zx}^\infty + 2F_{zy}^\infty\right) = \frac{a}{4}\left(3\ln(\sqrt{r^2+1}+r) + 3r\ln\left(\frac{\sqrt{r^2+1}+1}{r}\right) + \frac{1+r^3-(1+r^2)^{3/2}}{r}\right) \quad (A5)$$

In the currently interesting case of the finite rectangular rod, the Casimir limit can be further simplified by symmetry as $L_C = \frac{3}{2\pi}(F_{zx} + F_{zy} + F_{zz})$ by introducing the configuration factors associated to the three kinds of surfaces. $F_{zz}$ is in particular defined below with the width $a$ (along $x$), the thickness $b$ ($y$), the $z$-length $c$, with the notation $\alpha=2a/c$ and $\beta=2b/c$.

$$F_{zz} = \frac{c}{\alpha\beta}\left(\beta[\sqrt{\alpha^2+1}\arctan(\frac{\beta}{\sqrt{\alpha^2+1}}) - \arctan(\beta)] + \alpha[\sqrt{\beta^2+1}\arctan(\frac{\alpha}{\sqrt{\beta^2+1}}) - \arctan(\alpha)] + \frac{1}{2}\ln\left(\frac{(1+\alpha^2)(1+\beta^2)}{(1+\beta^2+\alpha^2)}\right)\right) \quad (A6)$$

The other configuration factors $F_{zx}$ and $F_{zy}$ are symmetric, and the former can be inferred from the latter according to Eq. (A7) with the replacement of $b$ by $a$ and vice versa. Since the last term in the following equation appears difficult to integrate analytically in contrast to the first two terms, $F_{zy}$ remains below as an integral form which can be numerically computed.

$$F_{zy} = \frac{2}{ab}\int_0^{c/2} dz \left(\frac{z^2}{2}\ln\left[\frac{z^2(a^2+b^2+z^2)}{(a^2+z^2)(b^2+z^2)}\right] + bz\arctan\left(\frac{b}{z}\right) - \frac{bz^2}{\sqrt{a^2+z^2}}\arctan\left(\frac{b}{\sqrt{a^2+z^2}}\right)\right) \quad (A7)$$

The use of Eq. (A6) and Eq. (A7) allows then to compute the Casimir limit for finite size samples.

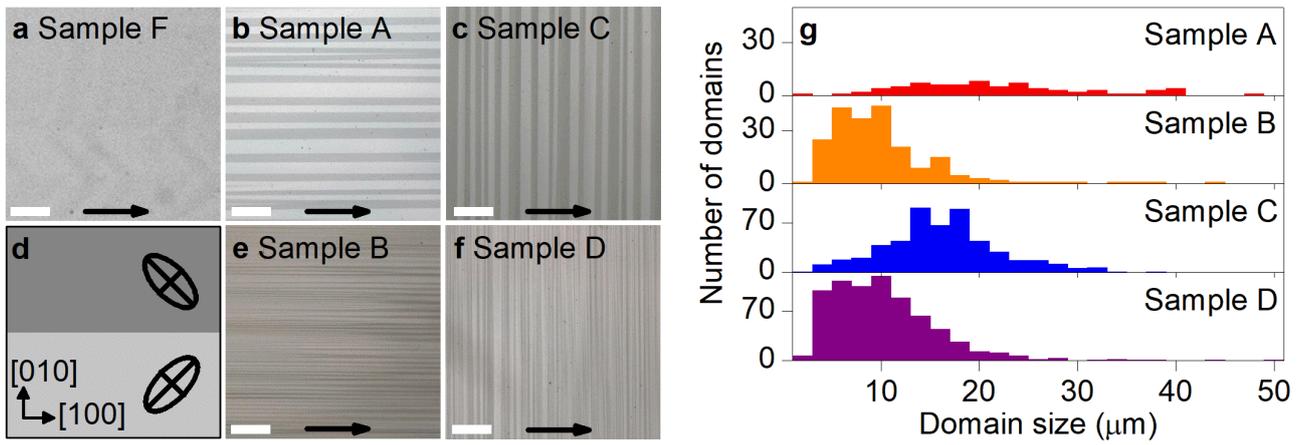

FIG. 1. Optical images in reflection of selected single crystals of LaAlO$_3$. (a) after annealing, (b,c) as-received, (e,f) after quenching. Panels are oriented such that the heat flow in thermal conductivity measurements goes from the left to the right, as indicated by black arrows.
(d) Schematics of the domain structure showing two domains, where ellipses indicate the orientation of their axes of compression (minor axis) and extension (major axis), projected into the plane of the schematic. Scale bars correspond to 100 μm. (g) Histogram of the number of domains with given sizes, measured from optical images.

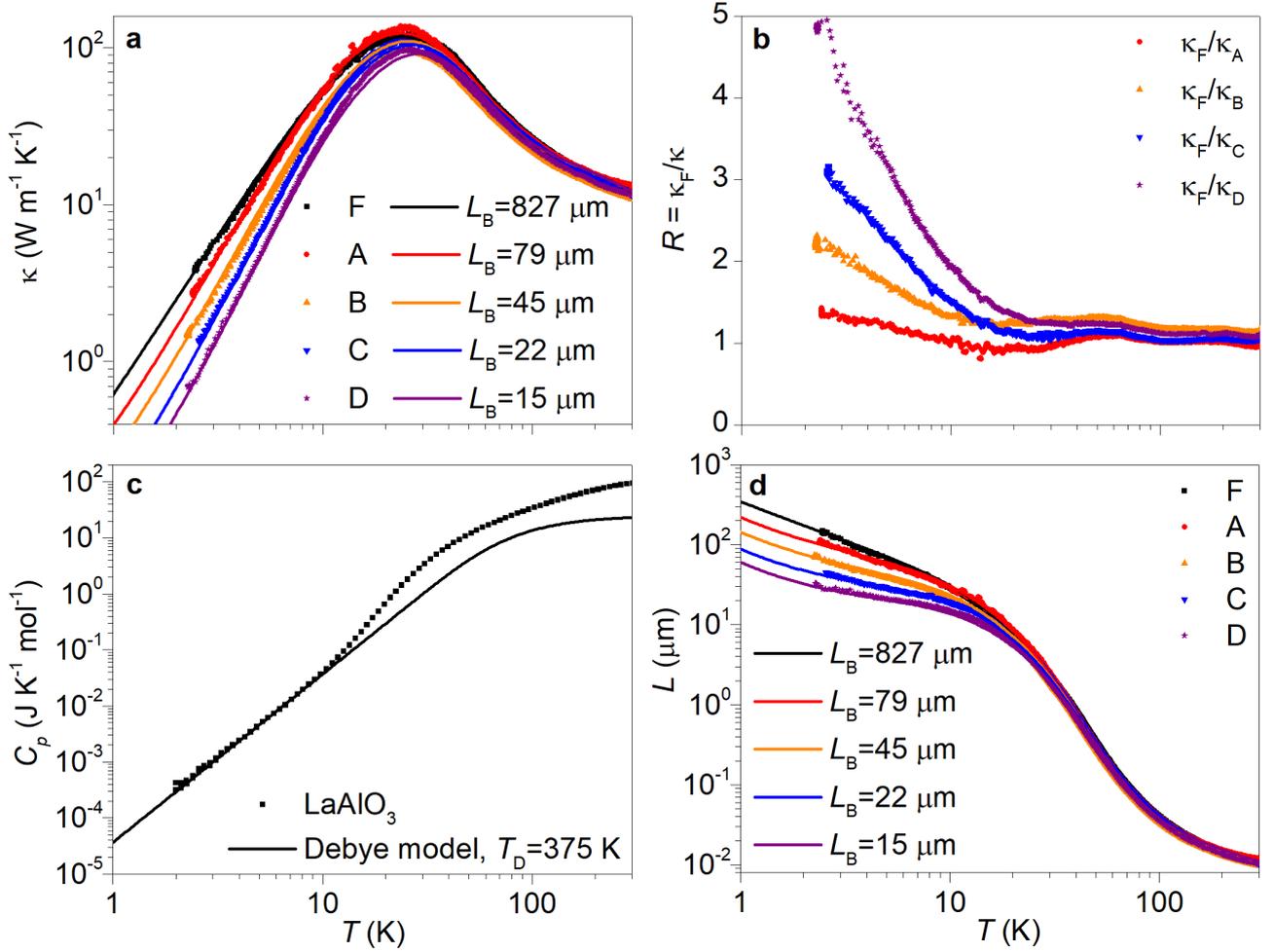

FIG. 2. Thermal conductivity as a function of temperature of selected LaAlO$_3$ samples displaying distinct domain wall densities and orientations. (a) Sample F is monodomain whereas in samples A and B domain walls are parallel to the heat flow, and orthogonal to it in C and D. Fitting lines in (a) originate from the Holland model including the corresponding boundary length $L_B$ as explained in the main text. (b) The switching ratio $R$ illustrates the enhancement of the monodomain thermal conductivity compared to the others. (c) Specific heat as a function of temperature measured in a single crystal of LaAlO$_3$. The line corresponds to the Debye model which accounts for the acoustic phonons' contribution. (d) Temperature dependence of the mean free path $L$ deduced from $\kappa = \frac{1}{3} C_{\text{vol}} v L$ with the inferred sound velocity $v$=4773 m s$^{-1}$ and $C_{\text{vol}} = C_{\text{p}} \left( \frac{M}{V M_{\text{mol}}} \right)$.

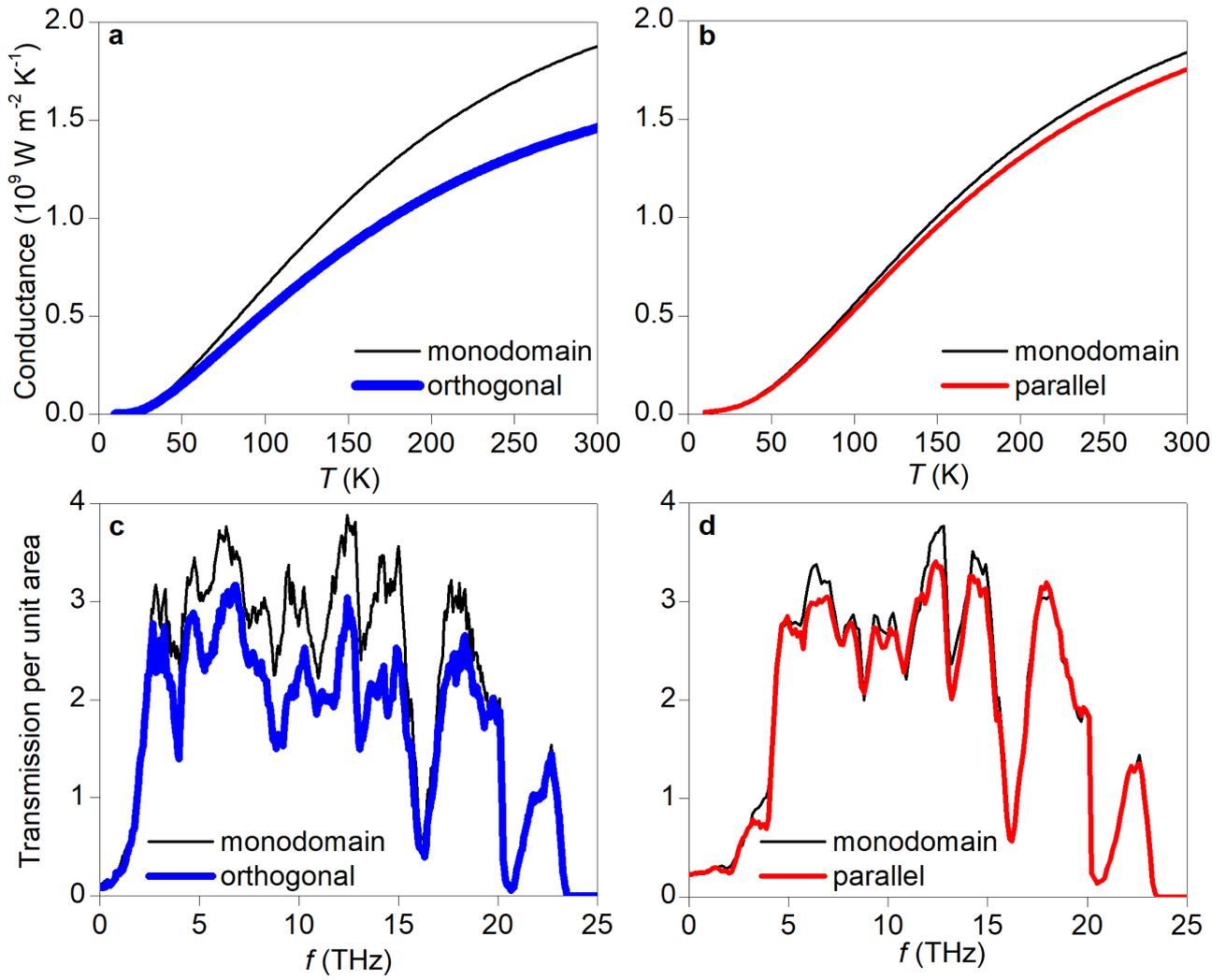

FIG. 3. (a,b) Thermal conductance as a function of temperature and (c,d) mode transmittance as a function of phonon frequency in the harmonic approximation, as simulated using non-equilibrium Green's function techniques. The black lines correspond to $LaAlO_3$ in the monodomain state, and blue (respectively red) lines in (a) and (c) [respectively (b) and (d)] correspond to $LaAlO_3$ with domain walls orthogonal (respectively parallel) to the heat flow.

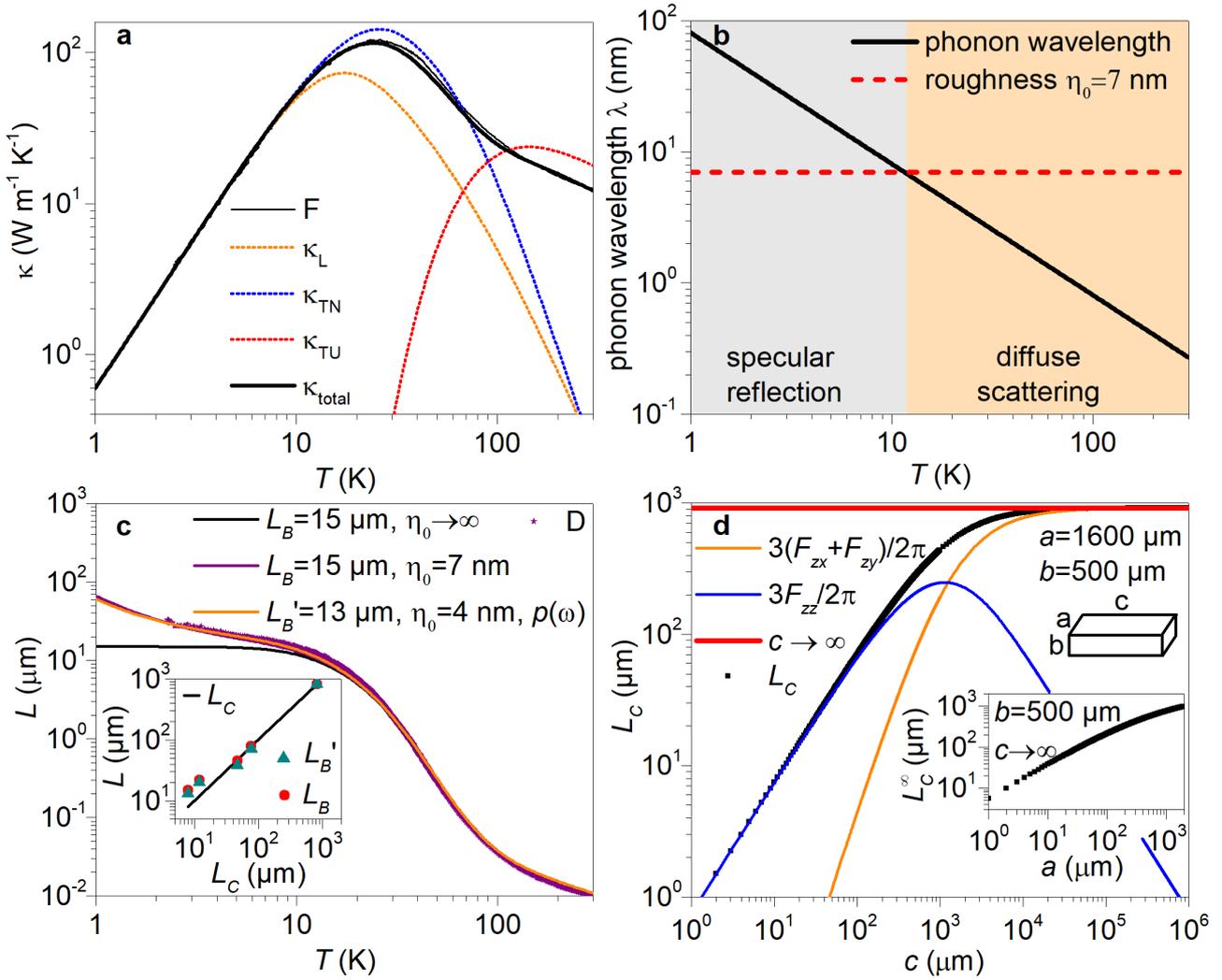

FIG. 4. Casimir limit and frequency dependent specularity parameter. (a) Temperature dependence of the thermal conductivity measured in the monodomain sample (F) compared to the calculated one according to the Holland model as $\kappa = \frac{1}{3}\kappa_L + \frac{2}{3}(\kappa_{TN} + \kappa_{TU})$. Dashed lines display the 3 contributions due to longitudinal and transverse modes, the latter resulting from Normal and Umklapp processes. The different scattering processes are summarized in Table 1. The boundary length used here is $L_B = 827$ μm. (b) Temperature dependence of the dominant phonon wavelength $\lambda \approx \frac{hv}{2.821 k_B T}$ compared to the rms roughness $\eta_0 = 7$ nm. The intersection of both curves defines a cross-over from a purely diffusive regime at high temperatures and a low temperature one where boundary reflections become specular. (c) It follows that the mean free path $L$ increases by exceeding the boundary length when the temperature is lowered, as for the sample D with its inferred mean free path $L$ compared to the calculated ones with and without specular reflection. The inset displays the variation of the boundary lengths including ($L_B'$) or not ($L_B$) a frequency dependent specularity parameter $p(\omega)$, as a function of the calculated Casimir limit related to the 5 investigated samples. (d) Variation of the Casimir limit as a function of $c$ according to Eqs. (A4), (A6) and (A7) for a finite rectangular rod compared to the result obtained with Eq. (A5) for an infinite one [59]. The inset displays the variation of the Casimir limit $L_C^\infty$ as a function of the width $a$ according to the latter equation.

| Scattering mechanism | Relaxation time $\tau^{-1}$ (s$^{-1}$) | Factor $A_i$ | Frequency range (s$^{-1}$) |
|---|---|---|---|
| Boundary | $\tau_B^{-1} = A_B$ | $A_B = v/L_B$ | $0 - \omega_D$ |
| Impurities (mass difference) | $\tau_I^{-1} = A_I \omega^4$ | $A_B = 3 \times 10^{-44}$ | $0 - \omega_D$ |
| Dislocations | $\tau_D^{-1} = A_D \omega$ | $A_D = 0; 1.15; 2.6 \times 10^{-5}$ | $0 - \omega_D$ |
| **Normal 3-phonon** | | | |
| longitudinal | $\tau_{NL}^{-1} = A_{NL} \omega^2 T^3$ | $A_{NL} = 1 \times 10^{-21}$ | $0 - \omega_D$ |
| transverse | $\tau_{NT}^{-1} = A_{NT} \omega T^4$ | $A_{NL} = 4 \times 10^{-11}$ | $0 < \omega_D/1.5$ |
| **Umklapp** | | | |
| longitudinal | $\tau_{UL}^{-1} = A_{UL} \omega^2 T^3 e^{-T_D/\alpha T}$ | $A_{UL} = 1.2 \times 10^{-22}, \alpha = 3$ | $0 - \omega_D$ |
| transverse | $\tau_{UT}^{-1} = \dfrac{A_{UT} \omega^2}{\sinh(\hbar\omega/k_B T)}$ | $A_{UT} = 9.3 \times 10^{-17}$ | $\omega_D/1.5 - \omega_D$ |

TABLE 1. List of scattering mechanisms of the Holland model extended with the dislocation term. The Debye frequency is deduced from the Debye temperature $T_D = 375$ K determined experimentally from the specific heat measurements as $\omega_D = k_B T_D/\hbar \approx 4.9 \times 10^{13}$ m s$^{-1}$. The inverse of the relaxation time of the boundary term is given by the ratio of the sound velocity $v \approx$ 4773 m s$^{-1}$ to the boundary length related to the Casimir limit $L_C$ as discussed in the main text.

| Samples | F | A | B | C | D |
|---|---|---|---|---|---|
| **Domain size (μm)** | monodomain | 19 | 9 | 15 | 9 |
| $L_B$ (μm) | 827 | 79 | 45 | 22 | 15 |
| $L_B'$ (μm) | 827 | 70 | 38 | 20 | 13 |
| $L_C$ (μm) | 827 | 78 | 47 | 12 | 8 |

TABLE 2. Boundary lengths. Average domain size measured on optical images. Boundary lengths with constant specularity parameter $p$ for $L_B$ ($\eta_0$=7 nm) or with a frequency dependent one $p(\omega)$ for $L_B'$ ($\eta_0$=4 nm). Casimir limit $L_C$ inferred from the optically determined domain distribution and Fig. 4(d).

[1] H. Hong, Y. H. Jung, J. S. Lee, C. Jeong, J. U. Kim, S. Lee, H. Ryu, H. Kim, Z. Ma, and T. Kim, *Anisotropic Thermal Conductive Composite by the Guided Assembly of Boron Nitride Nanosheets for Flexible and Stretchable Electronics*, Adv. Funct. Mater. **29**, 1902575 (2019).

[2] R. Ma, Z. Zhang, K. Tong, D. Huber, R. Kornbluh, Y. S. Ju, and Q. Pei, *Highly Efficient Electrocaloric Cooling with Electrostatic Actuation*, Science **357**, 1130 (2017).

[3] Y. Wang, Z. Zhang, T. Usui, M. Benedict, S. Hirose, J. Lee, J. Kalb, and D. Schwartz, *A High-Performance Solid-State Electrocaloric Cooling System*, Science **370**, 129 (2020).

[4] Y. Meng, Z. Zhang, H. Wu, R. Wu, J. Wu, H. Wang, and Q. Pei, *A Cascade Electrocaloric Cooling Device for Large Temperature Lift*, Nat. Energy **5**, 996 (2020).

[5] U. Ghoshal and A. Guha, *Efficient Switched Thermoelectric Refrigerators for Cold Storage Applications*, J. Electron. Mater. **38**, 1148 (2009).

[6] X. Gou, H. Ping, Q. Ou, H. Xiao, and S. Qing, *A Novel Thermoelectric Generation System with Thermal Switch*, Appl. Energy **160**, 843 (2015).

[7] M. Maldovan, *Sound and Heat Revolutions in Phononics*, Nature **503**, 209 (2013).

[8] N. Li, J. Ren, L. Wang, G. Zhang, P. Hänggi, and B. Li, *Colloquium: Phononics: Manipulating Heat Flow with Electronic Analogs and Beyond*, Rev. Mod. Phys. **84**, 1045 (2012).

[9] R. Xie, C. T. Bui, B. Varghese, Q. Zhang, C. H. Sow, B. Li, and J. T. L. Thong, *An Electrically Tuned Solid-State Thermal Memory Based on Metal-Insulator Transition of Single-Crystalline $VO_2$ Nanobeams*, Adv. Funct. Mater. **21**, 1602 (2011).

[10] S. Lee, K. Hippalgaonkar, F. Yang, J. Hong, C. Ko, J. Suh, K. Liu, K. Wang, J. Urban, X. Zhang et al., *Anomalously Low Electronic Thermal Conductivity in Metallic Vanadium Dioxide*, Science **355**, 371 (2017).

[11] J. Lee, E. Bozorg-Grayeli, S. Kim, M. Asheghi, H.-S. Philip Wong, and K. E. Goodson, *Phonon and Electron Transport through $Ge_2Sb_2Te_5$ Films and Interfaces Bounded by Metals*, Appl. Phys. Lett. **102**, 191911 (2013).

[12] J. Kimling, R. B. Wilson, K. Rott, J. Kimling, G. Reiss, and D. G. Cahill, *Spin-Dependent Thermal Transport Perpendicular to the Planes of Co/Cu Multilayers*, Phys. Rev. B **91**, 144405 (2015).

[13] J. A. Seijas-Bellido, C. Escorihuela-Sayalero, M. Royo, M. P. Ljungberg, J. C. Wojdeł, J. Íñiguez, and R. Rurali, *A Phononic Switch Based on Ferroelectric Domain Walls*, Phys. Rev. B **96**, 140101 (2017).

[14] S. Li, X. Ding, J. Ren, X. Moya, J. Li, J. Sun, and E. K. H. Salje, *Strain-Controlled Thermal Conductivity in Ferroic Twinned Films*, Sci. Rep. **4**, 6375 (2014).

[15] C. Liu, Y. Chen, and C. Dames, *Electric-Field-Controlled Thermal Switch in Ferroelectric Materials Using First-Principles Calculations and Domain-Wall Engineering*, Phys. Rev. Appl. **11**, 044002 (2019).

[16] C. Liu, P. Lu, Z. Gu, J. Yang, and Y. Chen, *Bidirectional Tuning of Thermal Conductivity in Ferroelectric Materials Using E-Controlled Hysteresis Characteristic Property*, J. Phys. Chem. C **124**, 26144 (2020).

[17] G. F. Nataf, M. Guennou, J. M. Gregg, D. Meier, J. Hlinka, E. K. H. Salje, and J. Kreisel, *Domain-Wall Engineering and Topological Defects in Ferroelectric and Ferroelastic Materials*, Nat. Rev. Phys. **2**, 634 (2020).

[18] B. Casals, G. F. Nataf, and E. K. H. Salje, *Avalanche Criticality during Ferroelectric/Ferroelastic Switching*, Nat. Commun. **12**, 345 (2021).

[19] R. J. Harrison and E. K. H. Salje, *Ferroic Switching, Avalanches, and the Larkin Length: Needle Domains in $LaAlO_3$*, Appl. Phys. Lett. **99**, 151915 (2011).

[20] A. J. H. Mante and J. Volger, *Phonon Transport in Barium Titanate*, Physica **52**, 577 (1971).

[21] W. Schnelle, R. Fischer, and E. Gmelin, *Specific Heat Capacity and Thermal Conductivity of $NdGaO_3$ and $LaAlO_3$ Single Crystals at Low Temperatures*, J. Phys. D. Appl. Phys. **34**, 846 (2001).

[22] M. A. Weilert, M. E. Msall, J. P. Wolfe, and A. C. Anderson, *Mode Dependent Scattering of Phonons by Domain Walls in Ferroelectric KDP*, Zeitschrift Fur Phys. B Condens. Matter **91**, 179 (1993).

[23] E. Langenberg, D. Saha, M. E. Holtz, J.-J. Wang, D. Bugallo, E. Ferreiro-Vila, H. Paik, I. Hanke, S. Ganschow, D. A. Muller et al., *Ferroelectric Domain Walls in PbTiO$_3$ Are Effective Regulators of Heat Flow at Room Temperature*, Nano Lett. **19**, 7901 (2019).

[24] J. F. Ihlefeld, B. M. Foley, D. A. Scrymgeour, J. R. Michael, B. B. McKenzie, D. L. Medlin, M. Wallace, S. Trolier-McKinstry, and P. E. Hopkins, *Room-Temperature Voltage Tunable Phonon Thermal Conductivity via Reconfigurable Interfaces in Ferroelectric Thin Films*, Nano Lett. **15**, 1791 (2015).

[25] B. M. Foley, M. Wallace, J. T. Gaskins, E. A. Paisley, R. L. Johnson-Wilke, J.-W. Kim, P. J. Ryan, S. Trolier-McKinstry, P. E. Hopkins, and J. F. Ihlefeld, *Voltage-Controlled Bistable Thermal Conductivity in Suspended Ferroelectric Thin-Film Membranes*, ACS Appl. Mater. Interfaces **10**, 25493 (2018).

[26] K. Aryana, J. A. Tomko, R. Gao, E. R. Hoglund, T. Mimura, S. Makarem, A. Salanova, M. S. Bin Hoque, T. W. Pfeifer, D. H. Olson et al., *Observation of Solid-State Bidirectional Thermal Conductivity Switching in Antiferroelectric Lead Zirconate (PbZrO$_3$)*, Nat. Commun. **13**, 1573 (2022).

[27] P. E. Hopkins, C. Adamo, L. Ye, B. D. Huey, S. R. Lee, D. G. Schlom, and J. F. Ihlefeld, *Effects of Coherent Ferroelastic Domain Walls on the Thermal Conductivity and Kapitza Conductance in Bismuth Ferrite*, Appl. Phys. Lett. **102**, 121903 (2013).

[28] S. Ning, S. C. Huberman, C. Zhang, Z. Zhang, G. Chen, and C. A. Ross, *Dependence of the Thermal Conductivity of BiFeO$_3$ Thin Films on Polarization and Structure*, Phys. Rev. Appl. **8**, 054049 (2017).

[29] J. Chrosch and E. K. H. Salje, *Temperature Dependence of the Domain Wall Width in LaAlO$_3$*, J. Appl. Phys. **85**, 722 (1999).

[30] M. A. Carpenter, A. Buckley, P. A. Taylor, and T. W. Darling, *Elastic Relaxations Associated with the Pm-3m –R-3c Transition in LaAlO$_3$ : III. Superattenuation of Acoustic Resonances*, J. Phys. Condens. Matter **22**, 035405 (2010).

[31] S. Kustov, I. Liubimova, and E. K. H. Salje, *LaAlO$_3$: A Substrate Material with Unusual Ferroelastic Properties*, Appl. Phys. Lett. **112**, 042902 (2018).

[32] See Supplemental Material at [URL will be inserted by publisher] for optical images in transmission, with crossed polarizer and analyzer, of the studied single crystals of LaAlO$_3$; analysis of ferroelastic domain structures; more details on thermal conductivity after quenching.

[33] X. Wang, U. Helmersson, J. Birch, and W.-X. Ni, *High Resolution X-Ray Diffraction Mapping Studies on the Domain Structure of LaAlO$_3$ Single Crystal Substrates and Its Influence on SrTiO$_3$ Film Growth*, J. Cryst. Growth **171**, 401 (1997).

[34] S. Bueble, K. Knorr, E. Brecht, and W. W. Schmahl, *Influence of the Ferroelastic Twin Domain Structure on the {100} Surface Morphology of LaAlO$_3$ HTSC Substrates*, Surf. Sci. **400**, 345 (1998).

[35] S. A. Hayward, S. A. T. Redfern, and E. K. H. Salje, *Order Parameter Saturation in LaAlO$_3$*, J. Physics. Condens. Matter **14**, 10131 (2002).

[36] J. J. R. Scott, B. Casals, K.-F. Luo, A. Haq, D. Mariotti, E. K. H. Salje, and M. Arredondo, *Avalanche Criticality in LaAlO$_3$ and the Effect of Aspect Ratio*, Sci. Rep. **12**, 14818 (2022).

[37] P. C. Michael, J. U. Trefny, and B. Yarar, *Thermal Transport Properties of Single Crystal Lanthanum Aluminate*, J. Appl. Phys. **72**, 107 (1992).

[38] J. Callaway, *Model for Lattice Thermal Conductivity at Low Temperatures*, Phys. Rev. **113**, 1046 (1959).

[39] P. G. Klemens, *The Scattering of Low-Frequency Lattice Waves by Static Imperfections*, Proc. Phys. Soc. Sect. A **68**, 1113 (1955).

[40] M. G. Holland, *Analysis of Lattice Thermal Conductivity*, Phys. Rev. **132**, 2461 (1963).

[41] E. Langenberg, E. Ferreiro-Vila, V. Leborán, A. O. Fumega, V. Pardo, and F. Rivadulla, *Analysis of the Temperature Dependence of the Thermal Conductivity of Insulating Single Crystal Oxides*, APL Mater. **4**, 104815 (2016).

[42] M. Moss, *Scattering of Phonons by Dislocations*, J. Appl. Phys. **37**, 4168 (1966).

[43] P. Carruthers, *Theory of Thermal Conductivity of Solids at Low Temperatures*, Rev. Mod. Phys. **33**, 92 (1961).

[44] H. B. G. Casimir, *Note on the Conduction of Heat in Crystals*, Physica **5**, 495 (1938).

[45] R. Berman, F. E. Simon, and J. M. Ziman, *The Thermal Conductivity of Diamond at Low Temperatures*, Proc. Roy. Soc. A **220**, 441 (1953).

[46] A. A. Maznev, *Boundary Scattering of Phonons: Specularity of a Randomly Rough Surface in the Small-Perturbation Limit*, Phys. Rev. B **91**, 134306 (2015).

[47] J. S. Heron, T. Fournier, N. Mingo, and O. Bourgeois, *Mesoscopic Size Effects on the Thermal Conductance of Silicon Nanowire*, Nano Lett. **10**, 2288 (2010).

[48] J. Cuffe, O. Ristow, E. Chávez, A. Shchepetov, P.-O. Chapuis, F. Alzina, M. Hettich, M. Prunnila, J. Ahopelto, T. Dekorsy, and C. M. Sotomayor Torres, *Lifetimes of Confined Acoustic Phonons in Ultrathin Silicon Membranes*, Phys. Rev. Lett. **110**, 095503 (2013).

[49] S. Hayward, F. Morrison, S. Redfern, E. Salje, J. Scott, K. Knight, S. Tarantino, A. Glazer, V. Shuvaeva, P. Daniel, M. Zhang and M. Carpenter, *Transformation Processes in $LaAlO_3$: Neutron Diffraction, Dielectric, Thermal, Optical, and Raman Studies*, Phys. Rev. B **72**, 054110 (2005).

[50] V. Janovec, *A Symmetry Approach to Domain Structures*, Ferroelectrics **12**, 43 (1976).

[51] V. Janovec and J. Přívratská, *Domain Structures*, in *International Tables for Crystallography* (International Union of Crystallography, Chester, England, 2006), pp. 449–505.

[52] J. C. Wojdeł, P. Hermet, M. P. Ljungberg, P. Ghosez, and J. Íñiguez, *First-Principles Model Potentials for Lattice-Dynamical Studies: General Methodology and Example of Application to Ferroic Perovskite Oxides*, J. Phys. Condens. Matter **25**, 305401 (2013).

[53] C. Escorihuela-Sayalero, J. C. Wojdeł, and J. Íñiguez, *Efficient Systematic Scheme to Construct Second-Principles Lattice Dynamical Models*, Phys. Rev. B **95**, 094115 (2017).

[54] G. Kresse and J. Furthmüller, *Efficient Iterative Schemes for Ab Initio Total-Energy Calculations Using a Plane-Wave Basis Set*, Phys. Rev. B **54**, 11169 (1996).

[55] G. Kresse and D. Joubert, *From Ultrasoft Pseudopotentials to the Projector Augmented-Wave Method*, Phys. Rev. B **59**, 1758 (1999).

[56] J. P. Perdew, A. Ruzsinszky, G. I. Csonka, O. A. Vydrov, G. E. Scuseria, L. A. Constantin, X. Zhou, and K. Burke, *Restoring the Density-Gradient Expansion for Exchange in Solids and Surfaces*, Phys. Rev. Lett. **100**, 136406 (2008).

[57] Z.-Y. Ong, *Tutorial: Concepts and Numerical Techniques for Modeling Individual Phonon Transmission at Interfaces*, J. Appl. Phys. **124**, 151101 (2018).

[58] A. Togo and I. Tanaka, *First Principles Phonon Calculations in Materials Science*, Scr. Mater. **108**, 1 (2015).

[59] J. P. Harrison and J. P. Pendrys, *Thermal Conductivity of Cerium Magnesium Nitrate*, Phys. Rev. B **7**, 3902 (1973).